\documentstyle[preprint,aps,epsfig,amstex,amssymb]{revtex}
\begin{document}
\draft
\title{How to break the replica symmetry in structural glasses}
\author{V.G.Rostiashvili$^{(a,b)}$, and T.A. 
  Vilgis$^{(a)}$} 
\address{$^{(a)}$Max-Planck-Institut f\"ur Polymerforschung,
  Postfach 3148, D-55021 Mainz, Germany} 
\address{$^{(b)}$
  Chemical Physics, Russian Academy of Science, 142432, Chernogolovka, Moscow
  region, Russia}
\maketitle
\begin{abstract}
The variational principle (VP) has been used to capture the metastable states
of a glass-forming molecular  system without  quenched disorder. It has been shown that VP naturally leads to a
self-consistent random field Ginzburg-Landau model (RFGLM). In the framework
of one-step replica symmetry breaking (1-RSB) the general solution of RFGLM is
discussed in the vicinity of the spinodal temperature $T_{A}$ in terms 
of ``hidden'' formfactors $\tilde g({\bf k})$,
$g_{0}({\bf k})$ and $\Delta({\bf k})$. The self-generated disorder
spontaneously arises.  It is argued
that at $T < T_{A}$ the activated dynamics is dominant.

\end{abstract}

%\setlength{\columnseprule}{0pt}
%\begin{multicols}{2}
It is a   common knowledge  that the transition from a liquid to
a glass is mainly determined by extremely slow dynamic processes. On  one
hand, the mode coupling theory (MCT) predicts a pure kinetic transition
temperature, $T_{c}$ , where the density fluctuations are frozen and the
system becomes nonergodic \cite{1}. On the other hand, there is a strong trend
in  favour of the idea that the appearance of metastable states in the free
energy landscape significantly determines the dynamical arrest at some
transition temperature $T_{A}$ \cite{2,3,4}. This consideration was based
 on both  Potts \cite{2,4} and $p$- spin \cite{3} glass models, in
 which  the
quenched disorder  was given. The authors  drew the conclusion
that these spin-glass models without reflection symmetry are also relevant for
the structural glasses, where the disorder in a sense should   not be quenched
but self-generated \cite{4'}.

Recently a great deal of work was done to incorporate the ideas and notions
of the mean-field theory of the spin-glasses  \cite{5} into the theory 
of the
structural glasses \cite{6,7,8,9,10}. The main  underlying
idea is related to the free-energy landscape. In order to quantify this landscape
paradigm  several copies ( or ``real replicas'' ) of the same system have been
introduced. In  doing so a reference replica $y$ is a typical configuration
distributed with a Boltzmann - Gibbs measure, whereas a replica $x$ is
affected by  the
influence of the replica $y$ through some short ranged attractive potential. It
is assumed that for a fixed value of $y$, which plays the role of the
quenched variables, the value of $x$ thermalized at the temperature of the
experiment. The small positive coupling constant  $\varepsilon $ is imposed
which forces the variables $x$ to stay near the $y$ variables. If this 
correlation between $x$ and $y$ variables persists even at
$\varepsilon \to 0$  the system is in a glassy phase. As a main result the expression for the ``effective potential''
has been calculated which shows  a nontrivial solution for the ``overlap''
$q(x,y)$ \cite{6,7,8}.

In this letter we shall present a different general method in order to treat the
metastable states in an arbitrary glass-forming molecular system  {\it
  without the 
quenched disorder}. For this purpose we reduce the initial problem to a
self-consistent random field Ginzburg-Landau model (RFGLM) by making use of the Feynman variational
principle (VP). We shall show that
the VP automatically yields  the standard replica structure in  phase space
with the final limit $n \to 0$ for the number of replicas. This RFGLM is
investigated by implementation of more  or  less standard field-theoretical
tools \cite{11,12,13,14}. The replica symmetry breaking (RSB) manifest 
itself as
a $1^{\rm st}$ order phase transition in the framework of a one-step RSB (1-RSB)
scenario and the self-generated disorder shows up. Some comparison
with the ``real replica'' method \cite{6,7,8,9,10} will be given below.

We shall start with a low-molecular system of $N$ particles with coordinates
${{\bf r}^{(p)}}$ , $(p = 1, ..., N)$ via a pair potential $V({\bf r})$. The
partition function at  temperature $T = 1/ \beta$ is given by 
\begin{eqnarray}
Z = \int \prod \limits_{p=1}^{N}d^{3}{\bf r}^{(p)}\exp
\left\{-\frac{\beta}{2}\sum \limits_{p,m=1}^{N}V({\bf r}^{(p)}- {\bf r}^{(m)}
  )\right\}. \label{1}
\end{eqnarray}

\noindent A physically  meaningful representation  of eq.~(\ref{1}) in
terms of  the  collective
density, $\rho ({\bf r}) = \sum_{p=1}^{N}\delta ({\bf r}- {
\bf r}^{(p)})$, is given by
\begin{eqnarray}
Z = \int D\rho (1)\exp \left\{W[\rho] - \frac{\beta}{2}\int d1d2 \rho (1)V(1 -
  2)\rho (2)\right\}, \label{2}
\end{eqnarray}

\noindent where the entropy of the  system is related to the Jacobian of the
corresponding transformation
\begin{eqnarray}
W[\rho] = \log \int \prod \limits_{p=1}^{N}d1^{(p)}\delta \left[\rho (1) -
  \sum \limits_{p=1}^{N} \delta (1 - 1^{(p)})\right] \label{3}
\end{eqnarray}

\noindent and the short - hand notation ${\bf r}^{(p)}\equiv 1^{(p)}$ has been used.
An alternative appropriate representation of eq.~(\ref{1}) can be obtained in terms
of a field $\psi (\bf r)$ conjugated to $\rho (\bf r)$  which appears as a
result of the functional Fourier transformation 
\begin{eqnarray}
\exp \left\{K[\psi]\right\} = \int D\rho (1)\exp \left\{W[\rho] - i \int d1
  \rho (1) \psi (1)\right\}, \label{4}
\end{eqnarray}

\noindent where $K[\psi]$ is the cumulant generating functional for the free system. The
functional expansion for $K[\psi]$ can be simply obtained in an explicit
form. The invertion of eq.~(\ref{4}) and substitution in eq.~(\ref{1}) gives
\begin{eqnarray}
Z = \int D\psi \exp \left\{K[\psi] - \frac{1}{2}\int d1d2\psi (1)[\beta
  V]^{-1}(1 - 2)\psi (2)\right\} . \label{5}
\end{eqnarray}

\noindent As a result,  eqs.~(\ref{2}) and (\ref{5}) provide two equivalent
representations of the partition function. 

Our main objective in the following  is to calculate the branch of the free energy which
appears in a supercooled (below the melting point) liquid and which  corresponds to
the metastable states. To this end we recall that in a glassy phase only the
component average free energy  
\begin{eqnarray}
\bar F = F_{c} + T\Sigma \label{6}
\end{eqnarray}
is physically meanigful \cite{15}. In eq.~(\ref{6}) the canonical free energy
$F_{c} = -T \log Z$ and the function $\Sigma \ge  0$ is called the complexity
\cite{15}. On the other hand it is known that the Feynman VP, when implemented for
the partition function (\ref{2}), yields an upper limit for  $F_{c}$ , 
i.e.$ F_{c} \le F_{\rm VP} $. As the {\it basic assumption} we take therefore

\begin{eqnarray}
\bar F = min \left\{ F_{0} +\left \langle \frac{\beta}{2}\int d1d2 \rho (1)V(1 -
  2)\rho (2) - W[\rho] - S[\rho]\right \rangle_{s}\right\} ,\label{7}
\end{eqnarray}

\noindent where the trial (Gaussian) Hamiltonian

\begin{eqnarray}
S[\rho] = \frac{1}{2}\int d1d2 \left\{G^{-1}(1,2)\rho (1)\rho(2)- 2
  G^{-1}(1,2)\rho(1)\langle \rho(2)\rangle_{s}\right\}. \label{8}
\end{eqnarray}

\noindent Here the expectation value $\langle \rho(1)\rangle_{s}$ and the correlator
$G(1,2) \equiv \langle \delta \rho (1)\delta \rho (2)\rangle_{s}$  are  considered as
independent variational functions. The assumption (\ref{7}) means that the
variational free energy $F_{\rm VP}$ takes its minimal value on the ensemble of components
(or pure states). It will be justified a posteriori by the appearance of the
replica structure. The minimization in eq.~(\ref{7}) with respect to $\langle
\rho(1)\rangle_{s}$ and $G(1,2)$  leads to 
\begin{eqnarray}
\langle \rho(1)\rangle_{s} = \int d2 [\beta V]^{-1}(1,2) \left \langle
  \frac{\delta}{\delta \rho (2)} W[\rho]\right\rangle_{s} , \label{9}
\end{eqnarray}
\begin{eqnarray}
G(1,2) = \left[ \beta V -\left \langle
  \frac{\delta^{2}}{\delta \rho (1)\delta \rho (2)}
  W[\rho]\right\rangle_{s}\right]^{-1}(1,2)\quad , \label{10}
\end{eqnarray}

where the ``quenched''$\psi$- expectation value reads

\begin{eqnarray}
\left \langle \frac{\delta}{\delta \rho (2)} W[\rho]\right\rangle_{s} = \left
  \langle \frac{\int D\psi i \psi (2)\exp \left\{ K[\psi] + i \int d1 \rho
      (1)\psi (1)\right\}}{\int D \psi \exp \left\{ K[\psi] + i \int d1 \rho
      (1)\psi (1)\right\} } \right \rangle_{s} \quad , \label{11}
\end{eqnarray}
and the ``quenched'' $\psi$ - correlator

\begin{eqnarray}
\left \langle \frac{\delta^{2}}{\delta \rho (1)\delta \rho (2)}
  W[\rho]\right\rangle_{s} = - \left
  \langle \frac{\int D\psi \delta\psi (1)\delta \psi (2)\exp \left\{ K[\psi] + i \int d1 \rho
      (1)\psi (1)\right\}}{\int D \psi \exp \left\{ K[\psi] + i \int d1 \rho
      (1)\psi (1)\right\} } \right \rangle_{s} \label{12}
\end{eqnarray}
 arises  naturally in this representation. As a result we have reduced the problem to a
self-consistent RFGLM . We have used the term ``quenched'' because  the Gaussian field $\rho(1)$ plays the role of an
``external quenched''  field in eqs.(\ref{11})- (\ref{12}).   Its
moments however 
should eventually  be determinated self-consistantly from
eqs.~(\ref{9})-(\ref{10}).

 One should not be  surprised that  Nature
reveals the replica structure through VP. It is well known (
see e.g. Sec.~4 in \cite{15'}) that the rigorous  treatment of
statistical thermodynamics yields only thermal equilibrium states. In
order to capture the metastable states in the framework of statistical 
thermodynamics one has to constrain properly the phase space of the
considered 
system. The most famous example is the van der Waals - Maxwell
metastable loop, which exists in the mean - field approximation and
becomes flatter if the clusters formation are taken into account
\cite{15'}.  It is the VP in our case
which implements this constraint. 

The
``quenched'' moments (\ref{11})-(\ref{12}) can be calculated by the standard
way \cite{5} from the {\it replicated partition function}  

\begin{eqnarray}
\left\langle Z^{n}\right\rangle_{s}\left\{G(1,2), \langle
  \rho(1)\rangle_{s}\right\} &=& \int \prod\limits_{a=1}^{n}D\psi_{a}\exp \Bigl  \{ - \frac{1}{2}\int d1d2 \sum \limits_{a,b=1}^{n}\left[ \rho_{0}\delta_{ab}
    \delta (1,2) + G(1,2)\right] \psi_{a}(1)\psi_{b}(2)\nonumber\\ &+& i \int d1 \langle \rho
    (1)\rangle_{s}\sum \limits_{a=1}^{n}\psi_{a}(1) + \sum
    \limits_{a=1}^{n}\tilde K[\psi_{a}]\Bigr\}\quad , \label{13}
\end{eqnarray}
where $\rho_{0}$ is an average density and $\tilde K[\psi_{a}]$ is the
anharmonic part of $K[\psi_{a}]$. As has been shown in ref.~\cite{11,12,13,14} the
effective Hamiltonian such as in eq.~(\ref{13}) may lead to the replica
symmetry breaking (RSB) which corresponds to a structural glass transition.

It is important to mention that for  potentials $V({\bf r})$ with the  infinite
range  interactions we can directly expand the effective Hamiltonian in
eq.~(\ref{2}) around the saddle point solution $\bar \rho (1)$ up to the
second order. The calculations show that this {\it next to the mean-field
  approximation} and the  VP merge  and that  both become exact,
i.e. $\bar F = F_{c}$. In this case the glassy phase does not appear. This conclusion
for the  present realistic model (\ref{1}) is in agreement with the result for a  ${\cal
  O}(M)$ -  model  with a quartic interaction in the large $M$ limit \cite{6} as well as for the
particles on a hypersphere \cite{16}. For the model given by eq.~(\ref{1})
with the infinite range of interaction the self-generated disorder is not
generic and  only crystallization  \cite{17} can be expected. It can
be shown \cite{17'} also that the Langevin dynamics of the
pure ( i. e. without a quenched disorder )  system
with the screened Coulomb potential $V({\bf r}) = (\mu /N)\exp
(-\gamma r)/r$  ( where the interaction parameter $\mu > 0$ , the
inverse screening length  $\gamma
\propto N^{-1/2}$ and   the number of particles $N >> 1$ ) does not
reveal any  glass transition. Despite
ongoing  discussions \cite{18} the glassy dynamic behavior of the pure
models is still debated.

Now we shall mention briefly the technical tools to calculate the free energy $
{\cal F} \left \{{\cal G}_{ab}(1,2), \left\langle \psi_{a}(1) \right\rangle
\right\}$ as a functional of the expectation value $\left\langle \psi_{a}(1)
\right\rangle $ and the correlator ${\cal G}_{ab}(1,2)$ for the replicated
field theory given by eq.~(\ref{13}). To this end, as in ref. \cite{13,14}, we have
used the second Legendre transformation. The detailed analysis shows that the
Hartree-like contribution into the so-called generating functional of all
2-irreducible diagrams is zero. The first nontrivial contribution comes from
the second-order terms in powers of vertices in eq.~(\ref{13}). The resulting
free-energy functional ${\cal F}$ has been parametrized in the framework of
the Parisi Anzatz \cite{5}:\quad  ${\cal G}_{ab}(1,2) \longrightarrow \left\{
  \tilde g({\bf k}), g({\bf k}, x)\right\}$ where the formfactor $\tilde
g({\bf k})$ corresponds to the diagonal and $g({\bf k},x)$ to the off-diagonal
elements of the hierarchical matrix ${\cal G}_{ab}(1,2)$. The wave vector {\bf
  k}- dependence is retained in the  spirit of  ref.\cite{19}. In
\cite{19} one can also find the beneficial formulas regarding the algebra of
the hierarchical matrices. In the off-diagonal formfactor $g({\bf k},x)$ the
argument $x \in  [0,1] $.

We have restricted ourselves to the simplest case of  1-RSB scenario \cite{5}. This means that the off-diagonal
formfactor $g({\bf k},x)$ consists only of  two pieces:  $g({\bf k},x) =
g_{0}({\bf k})$ at $x < x_{c}$ and  $g({\bf k},x) = g_{1}({\bf k})$ at $x >
x_{c}$ , where $x_{c}$ is the break point. It has been shown before that for
both Potts \cite{2,4} and $p$-spin glass models \cite{3,20} the 1-RSB scenario
is generic.

It is natural to define the gap $\Delta({\bf k}) \equiv g_{1}({\bf k}) -
g_{0}({\bf k})$ as a non-ergodicity (order) parameter of the required glassy
phase transition. As a consequence of these calculations  the
free-energy functional ${\cal F}\left\{\tilde g({\bf k}),g_{0}({\bf
    k}),\Delta({\bf k})\right\}$ has been obtained. The extremization of it with respect to
$\tilde g({\bf k}) $ and $ g_{0}({\bf k})$ at $\Delta({\bf k})= 0 $ yields
the equations for the replica symmetric (RS) case
\begin{eqnarray}
\frac{1}{\tilde g({\bf k}) - C({\bf k})} - \frac{1}{\tilde g({\bf k}) -
  g_{0}({\bf k}) } + \frac{ g_{0}({\bf k})}{[\tilde g({\bf k}) -
  g_{0}({\bf k})]^{2} } + \rho_{0} \nonumber\\
+ \frac{\rho_{0}^{2}}{2}\int \limits_{{\bf k}_{1}}\tilde g({\bf k}- {\bf
  k}_{1})\tilde g({\bf k}_{1}) - \frac{\rho_{0}^{2}}{6}\int \limits_{{\bf
    k}_{1}, {\bf k}_{2}}\tilde g({\bf k}- {\bf k}_{1} - {\bf k}_{2})\tilde
g({\bf k}_{1})\tilde g({\bf k}_{2}) = 0 \quad ,\nonumber\\
\frac{ g_{0}({\bf k})}{[\tilde g({\bf k})-g_{0}({\bf k})]^{2}}  +
\frac{1}{\tilde g({\bf k}) - C({\bf k})} + \frac{\rho_{0}^{2}}{2}\int \limits_{{\bf k}_{1}} g_{0}({\bf k}- {\bf
  k}_{1}) g_{0}({\bf k}_{1})\nonumber\\ 
- \frac{\rho_{0}^{2}}{6}\int \limits_{{\bf
    k}_{1}, {\bf k}_{2}} g_{0}({\bf k}- {\bf k}_{1} - {\bf k}_{2})
g_{0}({\bf k}_{1}) g_{0}({\bf k}_{2}) = 0 \quad ,\label{14}
\end{eqnarray}
where $\int\limits_{{\bf k}} \equiv \int d^{d} {\bf k}/(2\pi)^{d}$ and we
used the approximation $- \beta V({\bf k}) = C({\bf k})$ ($C({\bf k})$ is the
direct correlation function \cite{21}). For simplicity we have consided the
homogeneous case, $\left\langle \rho (1)\right\rangle_{s}= const $, to eliminate the
linear term with respect to $\psi$  in eq.~(\ref{13}). 

In order to grasp the glassy (or 1-RSB) solution , ${\cal
  F}\left\{\tilde g, g_{0},\Delta \right\}$ has been expanded up to the 4-th order with respect
to $\Delta ({\bf k})$ around the RS-solution (\ref{14}). As a result the
increment of the free energy connected with the non-zero order parameter
$\Delta ({\bf k})$ is determined by the Landau expansion
\begin{eqnarray}
{\cal F}_{1}\left\{\Delta \right\} 
=  -\int\limits_{{\bf q}_{1},{\bf q}_{2}}\Gamma_{2}\left({\bf q}_{1},{\bf
    q}_{2}; x_{c}\right)\Delta ({\bf q}_{1})\Delta ({\bf q}_{2}) -
\int\limits_{{\bf q}_{1},{\bf q}_{2}, {
\bf q}_{3}}\Gamma_{3}\left({\bf q}_{1},{\bf  q}_{2}, {\bf q}_{3};
x_{c}\right)\Delta ({\bf q}_{1})\Delta ({\bf q}_{2})\Delta ({\bf
q}_{3})\nonumber\\
-\int\limits_{{\bf q}_{1},{\bf q}_{2}, {
\bf q}_{3}, {\bf q}_{4}}\Gamma_{4}\left({\bf q}_{1},{\bf  q}_{2}, {\bf q}_{3},
{\bf q}_{4};
x_{c}\right)\Delta ({\bf q}_{1})\Delta ({\bf q}_{2})\Delta ({\bf
q}_{3})\Delta ({\bf q}_{4}) + ... \label{15}
\end{eqnarray}

The form of the expansion coefficients is determined by the RS-solution of
eq.~(\ref{14}) as it will be given in an extended paper \cite{22}. Since
$\Gamma_{3}$ is generally speaking nonzero the transition is
discontinuous. As  in ref.\cite{2,3,4}  our free energy is
proportional to the self-overlapping fraction :\quad ${\cal F}_{1} \propto - ( 1
- x_{c})$. In this case, as  has been argued in \cite{2,3,4}, the
metastable states appear  first at the spinodal point $T_{A}$ which is
determined by
\begin{eqnarray} 
\lim\limits_{x_{c}\to 1} \frac{\delta}{\delta\Delta ({\bf
    q})}\left\{\frac{{\cal F}_{1}\{\Delta \}}{1 - x_{c}}\right\}_{\Delta =
  \bar\Delta} = 0 \quad ,\label{16}
\end{eqnarray}
provided  that the eigenvalues of the matrix
\begin{eqnarray}
\chi^{-1}\left({\bf q}_{1},{\bf q}_{2}\right) = - \lim\limits_{x_{c}\to 1}\frac{\delta_{2}}{\delta\Delta ({\bf
    q}_{1})\delta\Delta ({\bf q}_{2})}\left\{\frac{{\cal F}_{1}\{\Delta \}}{1 - x_{c}}\right\}_{\Delta =
  \bar\Delta}  \label{17}
\end{eqnarray}
are nonnegative.

It is an open question whether the MCT transition temperature $T_{c}$ is related to $T_{A}$ . For  spin models with  quenched disorder  has been
shown that these temperatures  coincide \cite{2,3,4}. In any case at $T < T_{A}$ the
dynamics is dominated by the  activated process \cite{2,23}.

It is  interesting  that within  the ``hidden'' correlators $\tilde g({\bf k})$,
$g_{0}({\bf k})$ and $g_{1}({\bf k})$ only $\tilde g({\bf k})$ is related to
the density correlator:\quad $ G({\bf k}) = \left[\tilde g({\bf k}) - C({\bf
    k})\right]^{-1}$ (see eq.~(\ref{10})), so that the discontinuous transition does not shows up through
  the density correlator $G({\bf k})$ which is measured by an experiment. Nevertheless one can expect that below
  the spinodal point $T_{A}$ the activated dynamics becomes dominant and
  manifests itself in the time-dependent density correlator $ G({\bf k},
  t)$. We shall return to this question in the near future.
The equations for the RS-case (\ref{14}) should be solved numerically first
and then one can solve eq.(\ref{16})- (\ref{17}) for the order parameter $\bar\Delta({\bf k})$.

The glassy state appears first as a metastable one $(- {\cal
  F}_{1}\left\{\Delta \right\} \ge 0 ) $ but upon further cooling it can become
stable:\quad $- {\cal F}_{1}\left\{\Delta \right\} \le 0 $ (we recall that
the conditions (\ref{16})- (\ref{17}) determine the local maximum as usual in
spin - glasses \cite{5}). This possibility has been discussed in ref.\cite{24}
where on a basis of the Ramakrishnan - Yussouff free energy functional many
inhomogeneous ( glassy ) density configurations have been found. On the
contrary, in  spin - glasses the glassy state is thermodynamically
metastable for $T_{K}< T < T_{A} $ ( $T_{K}$ is the temperature  where the
complexity $ \Sigma$ becomes non-extensive)\cite{2,3,4,23}. 

We would shortly  like to  compare our method with the  mentioned ``real 
replica'' approaches \cite{6,7,8,9,10}. The main difference is that in
this method we deal with the conventional replicas where the ``zero
replica limit'' and RSB are  already fairly standard \cite{5}. This leads,
among other things, to a close link with the generalized spin-glass
models results \cite{2,3,4,24}. Our consideration based on the well
defined replicated field theory (see eq.~(\ref{13}) ) which needs as ``an
input'' only information about the direct correlation function
\cite{17}. Conversely, in the ``real replica'' approach one has to
choose one or other simple liquid models \cite{21} (e.g. the
replicated hypernetted chain (HNC) approximation in ref. \cite{8,9}
). Finally, our considerations can be directly generalized to the  Langevin
dynamics by making use of  the MSR- generating functional method
\cite{17',25}, whereas it is rather unclear what  a dynamical
counterpart of HNC in the simple liquid theory is.

In summary , we have suggested a general method to deal with the equilibrium
structural glass transition.  Making  use of the VP naturally leads to the
self-consistent RFGLM and the RSB - transition at some spinodal temperature
$T_{A}$. In the framework of 1-RSB scenario the equations for the set of
``hidden'' formfactors $\tilde g({\bf k})$,
$g_{0}({\bf k})$ and $\Delta({\bf k})$ are derived. Although the static
density correlator $G({\bf k})$ does not show  the RSB its dynamical
counterpart could be considerably affected by this transition through
the onset of the activated dynamics. 

The authors have benefited from discussions with J.Baschnagel,
K.Binder, W. Kob,  A.Heuer ,
K.M\"uller-Nedebock and  gratefully acknowledge financial support  by  the Sonderforschungsbereich 262.

%\end{multicols}
\end{document}